\documentclass[apj]{emulateapj}           
\slugcomment{Accepted by ApJ - Aug 23$^{rd}$ 2010} 
\usepackage{graphicx,amssymb,amsmath,times}
\usepackage{epsfig}
\usepackage[colorlinks=true,pdfstartview=FitV,linkcolor=blue,citecolor=blue,urlcolor=blue]{hyperref}

\def\Journal#1#2#3#4{{#4}, {#1}, {#2}, #3} 

\newcommand{\etal}{et al.}
\newcommand{\et}{\&}
\newcommand{\mysize}{\footnotesize}

\begin{document}

\title{Relative Composition and Energy Spectra \\ of Light Nuclei in Cosmic Rays: Results from AMS-01}
\shorttitle{Light Nuclei in Cosmic Rays}
\shortauthors{Aguilar \etal}

\author{
M.~Aguilar$^{y}$, 
J.~Alcaraz$^{y}$,
J.~Allaby$^{r}$\altaffilmark{$\dagger$}, 
B.~Alpat$^{ad}$, 
G.~Ambrosi$^{ad}$, 
H.~Anderhub$^{aj}$,
L.~Ao$^{g}$, 
A.~Arefiev$^{ab}$, 
L.~Arruda$^{x}$,
P.~Azzarello$^{ad}$
% ...
}


\affil{\mysize $^a$ I. Physikalisches Institut, RWTH, D-52074 Aachen, Germany\altaffilmark{3}} 
\affil{\mysize $^b$ III. Physikalisches Institut, RWTH, D-52074 Aachen, Germany\altaffilmark{3}}   
\affil{\mysize $^c$ LAPP, Universit\'e de Savoie, CNRS/IN2P3, F-74941 Annecy-le-Vieux Cedex, France}
\affil{\mysize $^d$ Johns Hopkins University, Baltimore, MD 21218, USA}
\affil{\mysize $^e$ Louisiana State University, Baton Rouge, LA 70803, USA}
\affil{\mysize $^f$ Center of Space Science and Application, Chinese Academy of Sciences, 100080 Beijing, China}
\affil{\mysize $^g$ Chinese Academy of Launching Vehicle Technology, CALT, 100076 Beijing, China}
\affil{\mysize $^h$ Institute of Electrical Engineering, IEE, Chinese Academy of Sciences, 100080 Beijing, China}
\affil{\mysize $^i$ Institute of High Energy Physics, IHEP, Chinese Academy of Sciences, 100039 Beijing, China\altaffilmark{4}} 
\affil{\mysize $^j$ University of Bologna and INFN-Sezione di Bologna, I-40126 Bologna, Italy\altaffilmark{5}}  
\affil{\mysize $^k$ Institute of Microtechnology, Politechnica University of Bucharest and University of Bucharest, R-76900 Bucharest, Romania}
\affil{\mysize $^l$ Massachusetts Institute of Technology, Cambridge, MA 02139, USA}
\affil{\mysize $^m$ National Central University, Chung-Li, Taiwan 32054}
\affil{\mysize $^n$ University of Maryland, College Park, MD 20742, USA}
\affil{\mysize $^o$ CHEP, Kyungpook National University, 702-701 Daegu, Korea}
\affil{\mysize $^p$ CNR--IROE, I-50125 Florence, Italy}
\affil{\mysize $^q$ Max--Planck Institut f\"ur extraterrestrische Physik, D-85740 Garching, Germany}
\affil{\mysize $^r$ European Laboratory for Particle Physics, CERN, CH-1211 Geneva 23, Switzerland}
\affil{\mysize $^s$ DPNC, Université de Genève, CH-1211 GENEVA 4 ,Switzerland}
\affil{\mysize $^t$ LPSC, Universit\'e Joseph Fourier Grenoble 1, CNRS/IN2P3, Institut Polytechnique de Grenoble, 38026 Grenoble, France}
\affil{\mysize $^u$ Chinese University of Science and Technology, USTC, Hefei, Anhui 230 029, China\altaffilmark{4}} 
\affil{\mysize $^v$ Helsinki University of Technology, FIN-02540 Kylmala, Finland}
\affil{\mysize $^w$ Instituto Superior T\'ecnico, IST, P-1096 Lisboa, Portugal}   
\affil{\mysize $^x$ Laboratorio de Instrumentacao e Fisica Experimental de Particulas, LIP, P-1000 Lisboa, Portugal}
\affil{\mysize $^y$ Centro de Investigaciones Energ{\'e}ticas, Medioambientales y Tecnol\'ogicas, CIEMAT, E-28040 Madrid, Spain\altaffilmark{6}} 
\affil{\mysize $^{z}$ INFN-Sezione di Milano, I-20133 Milan, Italy\altaffilmark{5}} 
\affil{\mysize $^{aa}$ Kurchatov Institute, Moscow, 123182 Russia}
\affil{\mysize $^{ab}$ Institute of Theoretical and Experimental Physics, ITEP, Moscow, 117259 Russia}
\affil{\mysize $^{ac}$ Universit\`a Degli Studi di Perugia, I-06100 Perugia, Italy} 
\affil{\mysize $^{ad}$ INFN-Sezione di Perugia, I-06100 Perugia, Italy\altaffilmark{5}}  

\affil{\mysize $^{ae}$ INFN-Sezione di Pisa and Universit\`a di Pisa, I-56100 Pisa, Italy\altaffilmark{5}} 

\affil{\mysize $^{af}$ INFN-Sezione di Roma, I-00185 Roma, Italy\altaffilmark{5}}  

\affil{\mysize $^{ag}$ Ewha Womens University, 120-750 Seoul, Korea}
\affil{\mysize $^{ah}$ Institute of Physics, Academia Sinica, Nankang Taipei 11529, Taiwan}
\affil{\mysize $^{ai}$ University of Turku, FIN-20014 Turku, Finland}
\affil{\mysize $^{aj}$ Eidgen\"ossische Technische Hochschule, ETH Z\"urich, CH-8093 Z\"urich, Switzerland}


\altaffiltext{$\bigstar$}{\mysize Corresponding author:~N.~Tomassetti (\href{mailto:Nicola.Tomassetti@pg.infn.it}{Nicola.Tomassetti@pg.infn.it})}

\altaffiltext{1}{\mysize Now at National Institute for High Energy Physics, NIKHEF, NL-1009 DB Amsterdam, The Netherlands.}

\altaffiltext{2}{\mysize Supported by ETH Z\"urich.}

\altaffiltext{3}{\mysize Supported by the 
Deutsches Zentrum f\"ur Luft-- und Raumfahrt, DLR.}

\altaffiltext{4}{\mysize Supported by the National Natural Science Foundation of China.}

\altaffiltext{5}{\mysize Also supported by the Italian Space Agency.}

\altaffiltext{6}{\mysize Also supported by the Comisi\'on Interministerial de Ciencia y Tecnolog{\'\i}a. }

\altaffiltext{$\dagger$}{\mysize Deceased.}


\begin{abstract}  
Measurement of the chemical and isotopic composition of cosmic rays is essential for the precise
understanding of their propagation in the galaxy.
While the model parameters are mainly determined using the B/C ratio,
the study of extended sets of ratios
can provide stronger constraints on the propagation models.
In this paper the relative abundances of the light nuclei lithium, beryllium, boron and carbon are presented.
The secondary to primary ratios Li/C, Be/C and B/C have been measured in the 
kinetic energy range $0.35 \-- 45\,$GeV~nucleon$^{-1}$. 
The isotopic ratio $^{7}$Li/$^{6}$Li is also determined in the magnetic rigidity interval $2.5 \-- 6.3\,$GV.
The secondary to secondary ratios Li/Be, Li/B and Be/B are also reported.
These measurements are based on the data collected by the Alpha Magnetic Spectrometer AMS-01 
during the STS-91 space shuttle flight in 1998 June. 
Our experimental results are in substantial agreement with other measurements, where they exist.
We describe our light-nuclei data with a diffusive-reacceleration model. 
A 10$\--$15$\,$\% overproduction of Be is found in the model predictions and can be
attributed to uncertainties in the production cross-section data. 
\end{abstract}


\section{Introduction}    
\label{sec::Introduction} 

The origin and properties of the charged cosmic rays (CRs) are
one of the major subjects of modern astrophysics.
Though experimental information comes from the analysis
of the arriving fluxes, the understanding of the relation between observational data 
and source properties requires a consistent picture of cosmic ray transport in the galaxy.
Propagation calculations take into account the acceleration, energy losses, nuclear interactions, 
magnetic diffusion and convective transport of CRs  
through the galactic medium~\citep{Strong2007,Maurin2001}.

The propagation of CR nuclei is studied using the ratio of secondaries,
which are created by fragmentation of heavier elements, to primaries, which are 
produced and accelerated in the astrophysical sources. 
The simple observation that the observed CR composition is different from that of rare solar system 
elements such as lithium, beryllium and boron proves the importance of propagation in the interstellar
medium (ISM) in terms of fragmentation processes.
In particular, the ratio B/C between boron and its main progenitor carbon is used
to constrain quantities such as the average amount of interstellar matter traversed by CRs between creation 
and observation, or their characteristic escape time from our galaxy.  
In descriptions based on diffusion theory, the secondary to primary ratios
are mainly sensitive to the energy dependence of the diffusion coefficient $D$.

The light elements Li and Be are also of interest. Their abundances depend not only on 
interactions of the primary species C, N and O, but also on tertiary contributions 
like Be$\rightarrow$Li or B$\rightarrow$Li. 
Therefore the Li/C and Be/C ratios may provide further restrictions on propagation
models~\citep{DeNolfo2006}.

An accurate understanding of CR properties is also of importance in the search for exotic signals 
in rare components of the cosmic radiation, as
the astrophysical background of any possible new physics signal
must be estimated on the basis of the existing models~\citep{Salati2010,Moskalenko2002}.

The origin and evolution of the elements Li-Be-B is also a crossing point between
different astrophysical fields: cosmology, astroparticle physics and nuclear physics~\citep{Reeves1994}.
The model of Big Bang Nucleosynthesis (BBN) is able to produce only faint traces of nuclei up
to $A=7$. Almost all the stars consume the relatively fragile Li-Be-B nuclei in their thermonuclear core reactions.
The galactic Li-Be-B are principally produced by the interaction of CRs with the ISM~\citep{Vangioni2000}.
Despite many observations and proposed solutions there are still open questions about the 
high value of the $^{7}$Li/$^{6}$Li ratio in meteorites 
and about the measured overabundance of primordial lithium~\citep{Asplund2006}. 

The CR chemical composition has been extensively studied
in a wide charge and energy range both on short duration balloon experiments
\citep{Webber1972, Orth1978, Lezniak1978, Simon1980, Buckley1994, RUNJOB1994}, 
long duration balloon flights~\citep{CREAM2008, ATIC2007}
and space experiments~\citep{Webber2002, HEAO1990, CRN1991}. 
Isotopic composition measurements come mainly from space experiments such as the 
High Energy Telescopes on {\it Voyager 1} and {\it 2} \citep{Webber2002} or the Cosmic Ray Isotope Spectrometer on the 
{\it Advanced Composition Explorer} satellite~\citep{DeNolfo2003}.
Measurements in a wider energy range, but with poorer statistics, have been obtained by several
balloon based magnetic spectrometers~\citep{Hams2004,Ahlen2000,WebberKish1979,Buffington1978}. 
A new generation of balloon borne experiments is devoted to extending the measured range toward 
the knee, where the power law decrease in the flux of CRs appears to steepen, by means of transition 
radiation detectors and calorimetric techniques~\citep{TRACER2009,CREAM2010,ATIC2007}.
The current and upcoming space experiments PAMELA~\citep{PAMELA2008} and 
the Alpha Magnetic Spectrometer \citep[AMS,][]{AMS01Report}, 
rely on magnetic spectrometers, in which
a magnetic field is used to bend the path of charged particles as they pass through. 
This allows accurate particle identification
together with a precise momentum determination free from atmospheric induced backgrounds.  

In this paper we present the complete analysis of our measurement of the CR charge composition 
and energy spectra in the range of $0.35\--45\,$GeV per nucleon.
Our preliminary results were presented for the B/C ratio in \citet{Tomassetti2009a}.
The $^{7}$Li to $^{6}$Li isotopic ratio is measured in the rigidity region 2.5$\,$GV to 6.3$\,$GV
(rigidity is defined as momentum per unit charge, $R=pc/Ze$).
These measurements take advantage of the large acceptance, accurate momentum resolution 
and good particle identification capabilities of the AMS-01 spectrometer.

\section{The AMS-01 Experiment} 
\label{sec::AMS01Detector}      

The Alpha Magnetic Spectrometer (AMS) is a particle physics detector designed for the 
high precision and long duration measurement of cosmic rays in space.
AMS-02 is scheduled to be installed on the {\it International Space Station} in 2010.
The AMS-01 precursor experiment operated successfully during a 10 day flight on
the space shuttle \textit{Discovery} (STS-91). 
Data taking started on 1998 June 3. The orbital inclination was $51.7^{\circ}$
and the geodetic altitude ranged from 320 to 390$\,$km.

The spectrometer was composed of a cylindrical permanent magnet, a silicon microstrip tracker,
Time-Of-Flight (TOF) scintillator planes, an aerogel \v{C}erenkov counter and anticoincidence counters~\citep{AMS01Report}.
The magnet (inner diameter 1.1$\,$m) provided a central dipole
field with an analyzing power $BL^2=0.14\,$Tm$^2$. Six layers of double sided silicon microstrip tracker
measured the trajectory of charged particles with an accuracy
of 10$\,\mu$m in the bending coordinate and 30$\,\mu$m in the nonbending coordinate, as well as providing multiple energy loss measurements.
The TOF system had two orthogonal segmented planes at each end of the magnet, and measured the particle
transit time with an accuracy of $\sim$90$\,$psec for $Z>1$ ions. 
The TOF scintillators also provided energy loss measurements up to $|Z|=2$.
A layer of anti-coincidence scintillation counters lined the inner surface of the magnet.
A thin carbon fiber layer was used as a shield to absorb low energy particles.

A total of 100 million triggers were recorded during the mission.
After the flight, the detector was calibrated
with helium and carbon beams at GSI, Darmstadt, and
with proton beams at the CERN PS, Geneva.
The data of the precursor flight have provided 
significant results on galactic and trapped protons, electrons,
positrons and helium nuclei~\citep{AMS01Report,FiandriniProton2004,FiandriniElectron2003}.
During the flight, nearly 200,000 nuclei with
charge $Z>2$ were recorded by AMS-01. 

\section{Data Analysis}   
\label{sec::DataAnalysis} 

The physical characteristics of a detected CR particle 
are the arrival direction, the particle identity and its momentum. 
These quantities were reconstructed combining the independent
measurements provided by the various detectors.
The particle rigidity was obtained by the deflection of the particle trajectory.
The velocity $\beta=v/c$ was determined from the transit time between
the TOF planes along the track length.
The charge magnitude $|Z|$ was obtained by the analysis of multiple 
measurements of energy deposition~(\S\ref{sec::ParticleIdentification}).

The response of the detector was simulated using the AMS simulation program, based on
\texttt{GEANT-3.21}~\citep{GEANT3} and interfaced with the hadronic package \texttt{RQMD}~\citep{RQMD}. 
The effects of energy loss, multiple scattering, interactions and decays were included, 
as well as detector efficiency and resolution.
Further details on reconstruction algorithms and detector performance 
are found in \citet{AMS01Report} 
and references therein.

During the STS-91 mission, the space shuttle \textit{Discovery} docked with the MIR
space station for about 4 days.  
The CR observations can be divided into four intervals:
\begin{enumerate}
\item[a)] \emph{pre-docking}, $\sim$1 day of testing and checkout before docking with MIR;
\item[b)] \emph{MIR-docked}, $\sim$4 days while the shuttle was docked with the MIR;
\item[c)] \emph{post-docking}, $\sim$3.5 days pointing at fixed directions 
($\psi=0^{\circ}$, $20^{\circ}$, $45^{\circ}$);
\item[d)] \emph{albedo}, $\sim$0.5 days pointing toward the Earth ($\psi=180^{\circ}$) .
\end{enumerate}
where $\psi$ is the angle between the AMS $z$-axis and the zenith.
For the present work we considered data collected during intervals b) and c).
We considered only particles traversing the detector top-down
within a restricted acceptance of $32^{\circ}$ from the positive $z$-axis.
Data collected while AMS-01 was passing in the region of the South Atlantic Anomaly
(latitude: $5^{\circ} \-- 45^{\circ} S$, longitude: $5^{\circ} \-- 85^{\circ}W$) were excluded.
\begin{figure}[!t]
\begin{center}
\epsscale{1.2}
\plotone {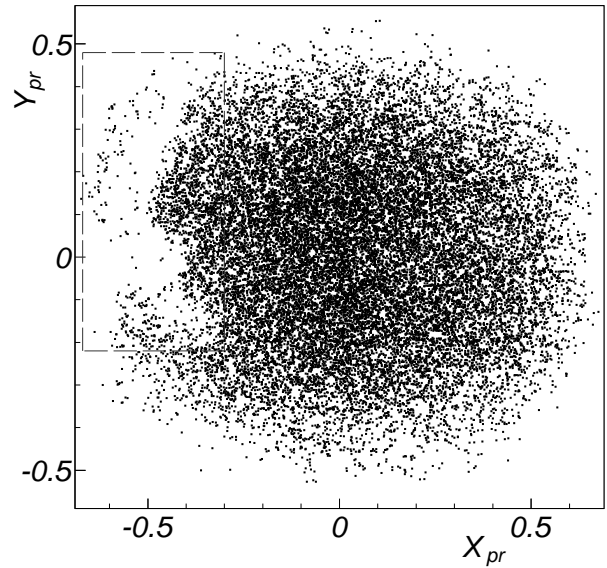}
\figcaption{
  Projection plot for downward going particles with measured charges $Z>2$ when 
  \textit{Discovery} was docked with MIR. Note the relative deficit of events from a clearly defined 
  region.  Candidates originating from that region were removed by a geometric cut on 
  the detector acceptance (dashed line).
  \label{fig::MirShadow}
}
\end{center}
\end{figure}
For data collected during b), a geometric cut on the detector
acceptance was imposed, corresponding to the 
MIR's ``shadow''.
In our previous work~\citep{Henning2005}, this region was recognized through the
secondary $\pi^{\pm}$ and $\mu^{\pm}$ produced by primary CRs interacting with the MIR modules.
In the nuclear channel ($Z>2$), a flux deficit can be observed in the same region. 
This is shown in Fig.~\ref{fig::MirShadow}, where the region above AMS-01 
is projected on to a $X_{pr}-Y_{pr}$ plane using the standard transformations of the arrival 
direction $X_{pr}=-\sin\theta \cos\phi$ and $Y_{pr}=\sin\theta \sin\phi$, 
where $\theta$ and $\phi$ are the polar and azimuthal angles in the AMS coordinate system of the incoming CR nuclei. 
Approximately 85,000 $Z>2$ candidates were selected with  
these criteria.

\subsection{Event Selection} 
\label{sec::DataSelection}   

In order to reject events with large nuclear scattering or poorly 
reconstructed trajectories, a set of quality cuts was applied to the selected candidates: 
\begin{figure*}[!t]
\begin{center}
\epsscale{0.95}
\plotone{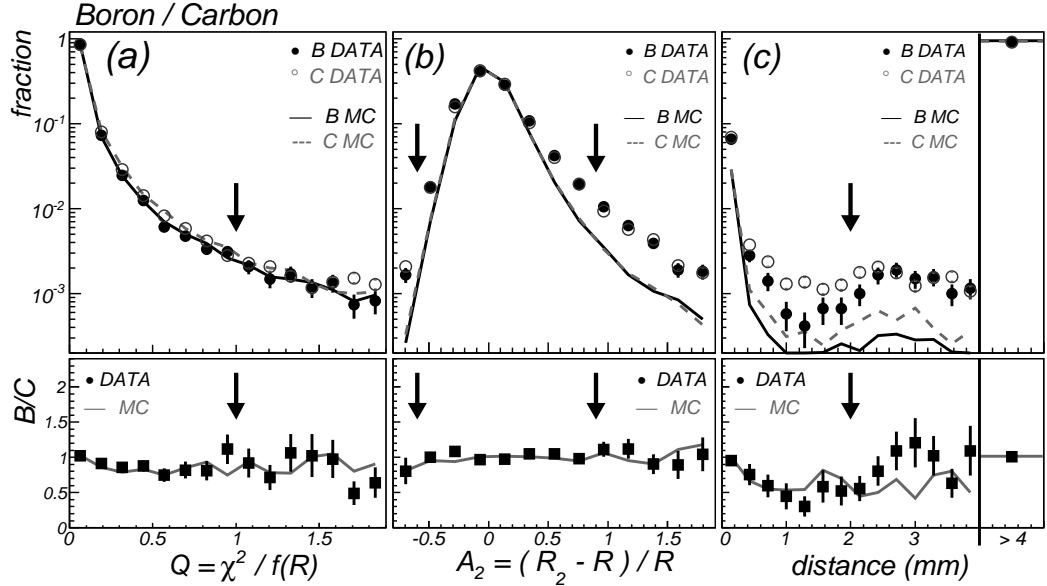} 
\figcaption{
      Top: normalized distributions of $Q= \chi^{2}/f(R)$ (a), $A_{2}$ (b),
      and unassociated hit distance (c) for detected boron (solid circles) and carbon (open circles) nuclei.
      In the cases that no hits were detected in the silicon or of the track extrapolation falling outside 
      of the active area, the distance is set in the $d>4\,$mm bin.  
      The same distributions from the MC simulation data are superimposed (lines).
      Bottom: ratios between the corresponding distributions of B and C are shown for data (squares) and MC (gray lines). 
      Cut thresholds are described by the arrows: events are accepted (a) to the left of the arrow, 
      (b) between the two arrows, and (c) to the right of the arrow.
      \label{fig::Cuts}
}
\end{center}
\end{figure*}
\begin{enumerate} 
\item tracks fitted with large ${\chi}^{2}$ were discarded according
   to a rigidity dependent requirement on $Q={\chi}^{2}/f(R)$, where $R$ is the measured rigidity;
 \item the particle rigidity  was measured as $R_{1}$ ($R_{2}$) using the first (last) three hits 
   along the reconstructed track.
   Tracks with poor agreement between $R$, $R_1$ and $R_2$ were removed according to cuts on 
   $A_{1}=\left( R_{1}-R \right)/R$ and $A_{2}=\left( R_{2}-R \right)/R$;
 \item consistency between $\beta$ and $R$ measurements was required. This cut acted on the
   tails of the reconstructed mass distributions of the detected particles.
 \item silicon planes with no hit associated with the track were inspected in the region 
   around the track extrapolation. 
   The absence of additional hits was required within a distance $d$ of 
   $2$ mm from the track extrapolation.
\end{enumerate}
As an indicator of mis-measured particles, a fraction of $\sim4\,$\% of events with
negative measured rigidity ($R<0$) was found in the reconstructed data before applying the selection cuts. 
These events were used as a control sample to define the selection.
The total selection efficiency was found to be substantially charge independent over 
the whole energy range considered.
In Fig.~\ref{fig::Cuts} the distributions of some quantities that were used for the selection
are shown for boron and carbon in comparison with Monte Carlo simulation (MC).
The agreement between data and MC in general is good; in the tails of the distributions, especially $A_2$,
the MC is less accurate in describing the measured behaviour (some few \% of the sample). 
However these features have similar magnitudes for the considered species and on average cancel out in the ratio,
as shown in the lower panels of the figure.
Similar conclusions can be drawn for the ratios Li/C and Be/C. 

\subsection{Particle Identification} 
\label{sec::ParticleIdentification}  

\begin{figure}[!t]
\begin{center}
\epsscale{1.22}
\plotone {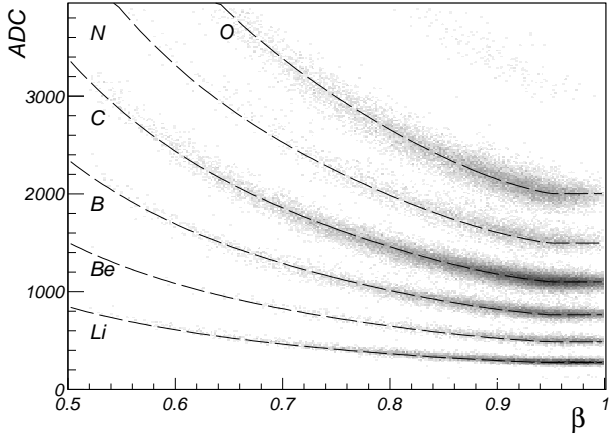} 
\figcaption{
  Mean energy loss in the silicon tracker as a function of the measured velocity $\beta$.
  The signal amplitude, shown in ADC count units, is the average signal of the best clusters $\{k_1,k_2,k_3\}$.
  Different nuclear species fall into distinct charge bands. 
  The $\beta$-$Z$ dependence of $\mu_{Z,\beta}$ is superimposed for $Z=3$ to $Z=8$ (dashed lines).
  \label{fig::dEdXvsBeta}
}
\end{center}
\end{figure}

Each elemental species is identified by its charge $Z$. 
The dynamical response of the TOF scintillators allowed particle discrimination up to $Z=2$.
The analysis of particle charge for $Z>2$ events was performed by the study of the energy losses recorded 
in the silicon layers.
The ionization energy generated by a charged particle in a silicon microstrip detector was
collected as a cluster of adjacent strips. Tracker clusters were recognized online
and then reprocessed with the reconstruction software.
A multistep procedure of normalization of the cluster signals was performed~\citep{Tomassetti2009b}.
The method accounted for saturation effects, electronics response, particle inclination and
velocity dependence of the energy loss.
The charge identification algorithm, applied to the corrected signals, was based on the maximum likelihood method
which determines the most likely value of $Z$ corresponding to the maximum value of the log-likelihood function:
\begin{equation}\label{EqLikelihood}
L(Z) = \log_{10}\{\prod_{k=1}^{k=6}P_{Z}^{k}(x_{k},\beta)\}
\end{equation}
The $k$-index refers to the six silicon layers, $x_{k}$ are the corrected
cluster signals as observed in the $k$-th layer.
The normalized $P_{Z}^{k}(x_{k},\beta)$ functions are the probabilities of a given charge
$Z$ with velocity $\beta$ to produce a signal $x_{k}$ in the $k$-th layer. These probability density
functions were estimated with a high purity reference sample of flight data, and describe
the energy loss distributions $x_{k}$ around a mean value $\mu_{Z,\beta} \propto (Z^2/\beta^2)\log(\gamma)$ 
as expected by the Bethe-Bloch formula.
This $\beta$-$Z$ dependence is shown in Fig.~\ref{fig::dEdXvsBeta} for $Z=3$ to $Z=8$ (dashed lines)
superimposed on the mean energy deposition from the measured tracker signals. 
Due to inefficiencies in charge collection,
some energy losses may produce charge responses which do not carry reliable
information on the particle charge. For this reason, not all
the six clusters were used to determine $Z$. 
For each reconstructed track, the most reliable set of three clusters
$\Omega=\{k_1,k_2,k_3\}$ was selected by also using 
$\Omega$ as parameter of log likelihood funtion: 
\begin{equation}\label{EqActualLikelihood}
L(Z,\Omega) = \log_{10}\{P_{Z}^{k_1}\cdot P_{Z}^{k_2}\cdot P_{Z}^{k_3} \}
\end{equation}
where the free parameters to be determined are $Z$, ranging from 3 (lithium) to 8 (oxygen),
and $\Omega$, running over all the 3-fold combinations $\{k_1,k_2,k_3\}$ of the tracker signals.
\begin{figure}[!t]
\begin{center}
\epsscale{1.13}
\plotone {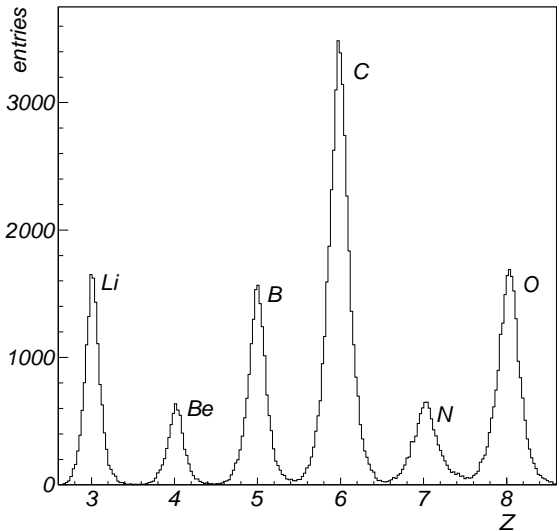}
\figcaption{
  Charge spectrum of the selected $Z>2$ data.  
  The signal amplitudes of Fig.~\ref{fig::dEdXvsBeta} are equalized to $\beta\equiv 1$
  and shown in units of charge.
  Different nuclear species fall in distinct charge peaks. 
  \label{fig::DEDX}
}
\end{center}
\end{figure}
Though the final estimation of $Z$ was provided by three tracker clusters,
all the selected clusters were processed in Eq.~\ref{EqActualLikelihood}.
An identification inefficiency of a few percent was achieved with this algorithm.
On the basis of the final charge spectrum shown in Fig.~\ref{fig::DEDX}, the determination of B and C 
was done with a charge resolution better than $0.14$ charge units.

\subsection{Charge Contamination} 
\label{sec::ChargeContamination}  

The charge assignment procedure was studied in terms of efficiency and contamination.
Each nucleus of charge $3\leq Z\leq 8$ produces a charge estimation $\hat{Z}$ which can  be
related to its \emph{true} impinging charge $Z$ by using a set of coefficients $F_{Z}^{\hat{Z}}$.

The 6$\times$6 matrix $\|F\|$ is diagonally-dominated, and each off-diagonal element
$F_{Z}^{\hat{Z}}$ represents the probability of a nuclear species $Z$ to be
misidentified as $\hat{Z}$ due to interactions in the detector material and fluctuations
of the energy loss:
\begin{equation}\label{EqMigrationProbability}
F_{Z}^{\hat{Z}}=P(\hat{Z}|Z)
\end{equation}
Two different contributions produce a charge migration $Z\rightarrow \hat{Z}$:
\begin{enumerate}
  \item[A)] After interacting in the upper TOF material, an incoming nucleus $Z$ may fragment and 
    physically turn into $\hat{Z}$.
    These events were typically rejected by the anticoincidence veto. However, a fraction of them produces
    a clean track in the tracker, passing trigger and selection. Since $\hat{Z} < Z$, the corresponding matrix is triangular.
  \item[B)] Fluctuations of the measured energy loss produce a nonzero $Z\rightarrow \hat{Z}$ migration probability.
    This charge spill over is typically symmetric and in most cases comes from adjacent charges, i.e. $\hat{Z}=Z\pm 1$.
\end{enumerate}
Both the effects were studied with Monte Carlo simulations. 
The second contribution was also estimated with the data.
Possible charge migrations from/to the $Z<3$ and $Z>8$ species were also considered, in particular
to achieve a suitable separation of the $Z>2$ candidates from the background of the more abundant helium flux. 
Helium-lithium separation was perfomed with the combination of both TOF and tracker energy depositions
and studied with the help of data from the calibration beam test during which
the detector was probed with pure helium beams at 0.75, 1.8, 3.6 and 8 GeV~nucleon$^{-1}$ of kinetic energy~\citep{AMS01Report}. 
Contamination from He to the selected $Z>2$ sample was estimated to be less than $10^{-4}$ of the helium sample.
The effects of charge misassignment in the measured secondary to primary ratios 
turned out to be smaller than the statistical uncertainties; 
this effect is included in the systematic errors~(\S\ref{sec::ErrorBreakdown}).

\section{Flux Determination} 
\label{sec::FluxCalculation} 

\begin{figure}[!t]
\begin{center}
\epsscale{1.08}
\plotone {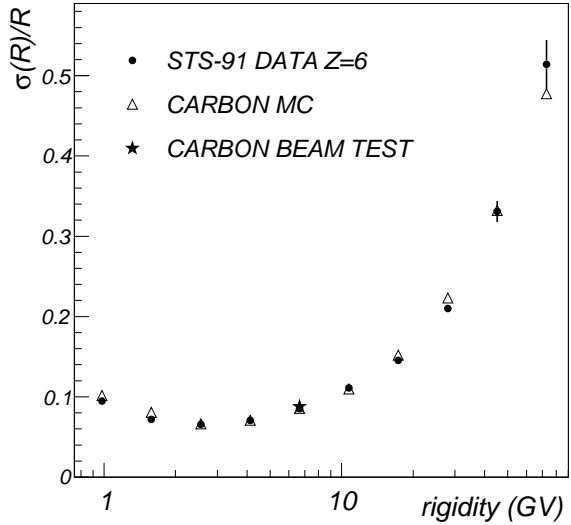}
\figcaption{
  Rigidity resolution of the AMS-01 silicon tracker estimated with measured data from the flight
  and the calibration test beam with carbon nuclei in comparison with the MC simulation.
\label{fig::ResolutionCarbon}
}
\end{center}
\end{figure}

Before the secondary to primary ratios can be determined the spectra must be transformed into fluxes $\Phi$.
The differential energy spectrum of the $Z$-charged particles measured by AMS-01 in
the energy bin $E$ of width $\Delta E$ is related to the measured counts $\Delta N^{Z}$ by:
\begin{equation}\label{EqFluxSimplified}
\Phi^{Z}(E)=\frac{\Delta N^{Z}(E)}{A^{Z}(E)\cdot \Delta T^{Z}(E)\cdot \Delta E}
\end{equation}
where $\Delta T^{Z}(E)$ is the effective exposure time (\S\ref{sec::ExposureTime})
and $A^{Z}(E)$ is the detector acceptance (\S\ref{sec::AcceptanceEstimation}).
The relation between the measured energy of detected particles and their true energy
were studied using unfolding techniques~\citep{DAgostini1995}.
Making use of an appropriate energy binning, the final smearing effects turned out to be barely appreciable 
and very similar for the variuous species, \textit{i.e.}, the overall effect is mostly canceled out in the bin-to-bin ratios. 
The rigidity resolution curve of AMS-01 is shown in Fig.~\ref{fig::ResolutionCarbon} for carbon nuclei
measured during the flight and in the beam test, together with the simulation results.

\subsection{Exposure Time} 
\label{sec::ExposureTime}  

By the term exposure time $\Delta T^{Z}(E)$, we mean the effective data taking duration with
AMS-01 capable of receiving a new trigger from a galactic particle of charge $Z$ in the energy interval $\Delta E$.
The detection is influenced by the DAQ livetime fraction $\alpha$ and the geomagnetic field modulation.
Though no atmospheric or trapped particles are expected in our $Z>2$ fluxes,
the position dependent geomagnetic cutoff introduces different distortions of the 
measured energy spectrum for different nuclear species.
Hence only particles with rigidity greater than $R^{Th}=1.25 \cdot R_{C}$ were accepted, 
where $R_{C}$ is the rigidity cutoff obtained from the St\"ormer estimation 
in the corrected geomagnetic coordinates for 1998~\citep{SmartShea2005}. 
The entire orbit was divided into 45,000 time intervals of $\delta t \sim10\,s$. 
For each time bin $\delta t_{k}$, the mean values of $R^{Th}_{k}$ and $\alpha_{k}$ (trigger livetime fraction) were computed.
The exposure time $\Delta T^{Z}(E)$ was then calculated for each particle $Z$, rigidity $R$
and energy $E \in \Delta E$ as:
\begin{equation}\label{EqExposureTime}
\Delta T^{Z}(E)= \sum_{k} \alpha_{k} \cdot H^{Z}_{k}(R,R_{k}^{Th}) \cdot \delta t_{k}
\end{equation}
where $R= \left( \frac{A}{Z} \right) \sqrt{E^{2}+ 2M_{p}E}$,
$H=1$ for $R>R^{Th}$, and $H=0$ for $R<R^{Th}$,
\textit{i.e.}, the sum is restricted to only the time intervals above cutoff. 
The result describes the effective exposure of AMS-01 to
cosmic rays in that energy bin coming from outside the magnetosphere. 
Note that $\Delta T^{Z}(E)$ as a function of the kinetic energy per nucleon is different for different $A/Z$ isotopes;
since $A/Z$ is slightly different for the elements under consideration, their corresponding exposure times do not completely 
cancel in the flux ratios. 

\subsection{Acceptance Estimation} 
\label{sec::AcceptanceEstimation}  
The detector acceptance $A^{Z}(E)$ of Eq.~\ref{EqFluxSimplified} is the convolution of geometrical factors
with the energy dependent efficiency $\epsilon$, including trigger, reconstruction and selection efficiencies.
The acceptances of each species $A^{Z}$ were determined with the Monte Carlo simulation~\citep{Sullivan1971}. 
2$\cdot 10^{8}$  trajectories were generated in the energy range $0.2\--50\,$GeV~nucleon$^{-1}$, 
according to an isotropic distribution and a power law momentum distribution $\propto p^{-1}$. 
The recorded MC events were sent through the same analysis chain as for the measured data.
Background fluxes were also simulated, up to $Z=8$.

\subsection{Systematic Errors} 
\label{sec::TriggerEfficiency} 
\label{sec::ErrorBreakdown}    

Though systematic uncertainties arising from many steps of the analysis are suppressed 
in the ratios, differences in the trigger efficiencies of the two species 
are present, as expected, from the charge dependence of delta ray production and 
fragmentation effects in the detector material. The spill over from adjacent charges also 
produces net effects on the measurements.

The MC determination of the acceptance gave a statistical uncertainty 
of $\sim$1$\--$4$\,$\%, increasing with energy. 
$A^{Z}(E)$ decreases with increasing energy and charge,
due to the trigger conditions against interacting particles. 
In the events collected from the flight, the dedicated unbiased dataset\footnote{
One out of 1000 events was recorded with the AMS-01 unbiased trigger.} did not give enough
statistics for an energy dependent validation of the $Z>2$ trigger efficiency with measured data,
hence this estimation must rely on simulation results.
Since CR interactions with the detector material can play an important
role in the observed behaviour of the trigger response,
a \texttt{GEANT3}-based MC simulation does not guarantee a 
model independent description of the experimental setup.
\begin{figure*}[!bhtp]
\begin{center}
\epsscale{0.85}
\plotone{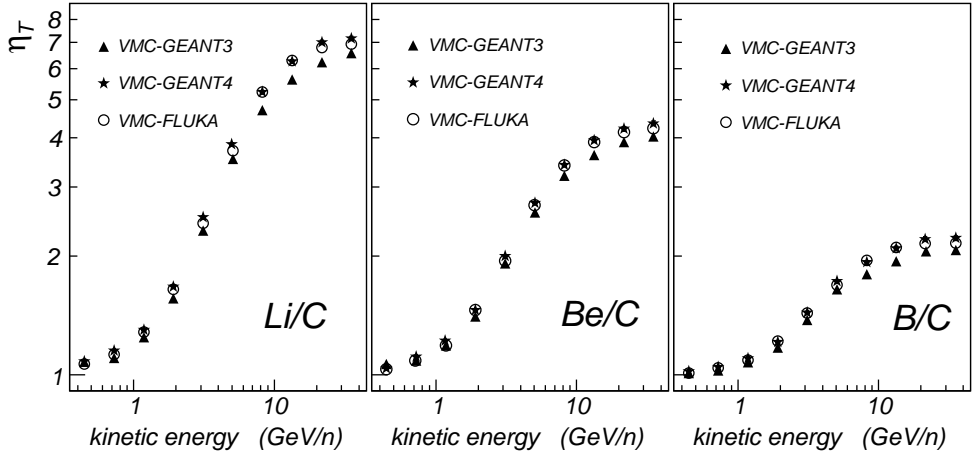} 
\figcaption{
  Ratio of acceptances $\eta_{T}$ estimated with \texttt{GEANT3}, \texttt{GEANT4}
  and \texttt{FLUKA} based simulations implemented under the \texttt{AMS-VMC} application.
  \label{fig::VMC}
}
\end{center}
\end{figure*}
Therefore, this study was performed with three different models.
A \texttt{Virtual MC} application~\citep{VMC} was developed for AMS-01. 
The simulation natively supports the particle transport codes \texttt{GEANT3} and \texttt{GEANT4}~\citep{GEANT4}.
An additional interface with \texttt{FLUKA}~\citep{FLUKA} was developed~\citep{Oliva2007, Tomassetti2009b}.
Since the quantities that enter in to our measurements are the ratios $\eta_{T}$
between the corresponding acceptances $A^{Li}/A^{C}$, $A^{Be}/A^{C}$ and $A^{B}/A^{C}$, 
the systematic errors on the trigger efficiency were estimated observing the scatter between 
the three different models.
Results of this
approach are summarized in Fig.~\ref{fig::VMC}. 
From the figure the charge dependent behaviour of the trigger response is clear; 
the energy dependence of the trigger efficiency is more pronounced for higher charges.
Good agreement was found between \texttt{GEANT4} and \texttt{FLUKA};
nevertheless, for each ratio, the full envelope of the three model curves was considered as
the uncertainty band of the trigger efficiency.

Another source of uncertainty comes from the isotopic composition of the measured nuclei.
In order to compute bin-to-bin ratios, all the elemental fluxes were determined in a common 
grid of kinetic energy per nucleon $E$.
The rigidity to energy conversion required the $A/Z$ ratio of each charged species. 
To first approximation $A/Z \approx 2$ for the light $Z>2$ nuclei under study; 
however, according to existing measurements and model calculations, more realistic isotopic mixtures 
can be considered for the Li-Be-B group~\citep{Hams2004,Ahlen2000,Webber2002,Buffington1978}.
Hence we assumed isotopic fractions
$^{6}$Li/($^{6}$Li+$^{7}$Li)~$= 0.5 \pm 0.15$, 
$^{7}$Be/($^{7}$Be+$^{9}$Be)~$= 0.65 \pm 0.15$, and 
$^{10}$B/($^{10}$B+$^{11}$B)~$= 0.35 \pm 0.15$,
accounting for these uncertainties in the estimation of the systematic error. 
\begin{figure*}[!t]
\begin{center}
\epsscale{0.75}
\plotone{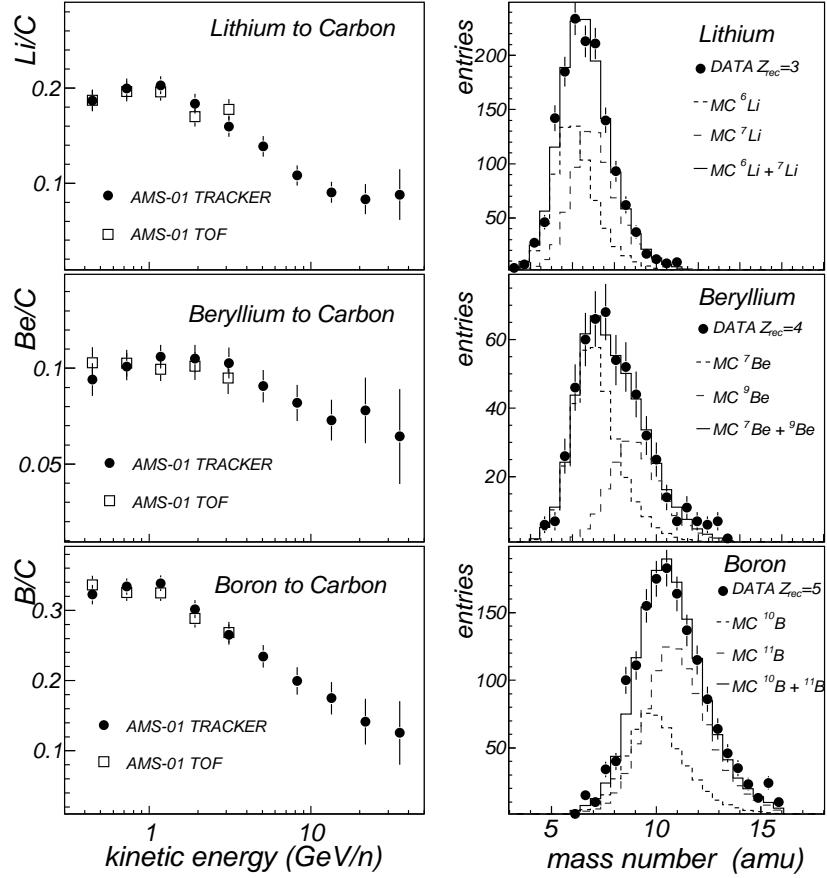} 
\figcaption{
  Left: consistency check of the Li/C, Be/C and B/C ratio obtained with tracker measurements between 0.35
  and 45$\,$GeV~nucleon$^{-1}$ (filled circles) in comparison with that measured with the TOF between $0.35$ 
  and 4$\,$GeV~nucleon$^{-1}$ (open squares). Error bars include both statistical and systematic errors.
  Right: mass number distribution of measured (filled circles) and 
  simulated (lines) isotopes for $\beta<0.95$. 
  MC curves are normalized to the data and contain isotopic mixtures of
  $^{6}$Li\--$^{7}$Li, $^{9}$Be\--$^{7}$Be and $^{10}$B\--$^{11}$B.
  \label{fig::MassComposition}
}
\end{center}
\end{figure*}
In the left panels of Fig.~\ref{fig::MassComposition} we cross checked our measured ratios with 
independent measurements obtained with the TOF system, which provided a direct measurement of 
the velocity and hence kinetic energy per nucleon. 
Results from the tracker measurement and the TOF measurement are consistent.
The right panels of Fig.~\ref{fig::MassComposition} show the reconstructed mass distributions 
for the Li-Be-B events above cutoff with $\beta<0.95$. 
Though the AMS-01 mass resolution did not allow an event-to-event isotopic separation, 
the data are in good agreement with the MC simulation which contains the assumed mixtures. 

\begin{figure}[!h]
\begin{center}
\epsscale{1.0}
\plotone{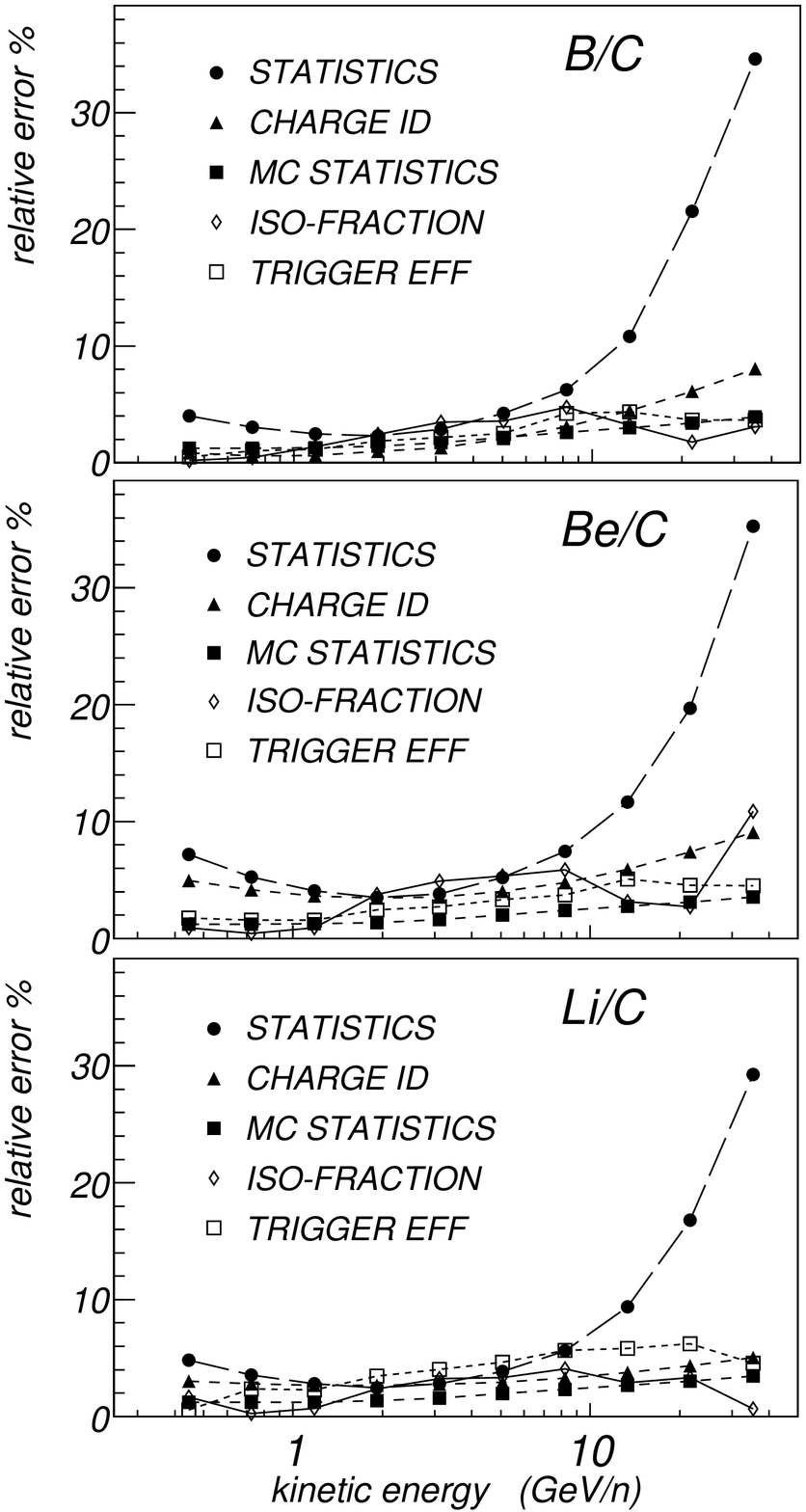} 
\figcaption{
  Relative errors on the measured ratios Li/C, Be/C and B/C.
  The contributions from data statistics, Monte Carlo statistics,
  contamination on the charge identification, isotopic composition and trigger efficiency are shown as a 
  function of the kinetic energy per nucleon. 
  For the sake of graphical clarity, points are connected with lines.
  \label{fig::ErrorBreakdown}
}
\end{center}
\end{figure}

A detailed summary of the uncertainties on the measured ratios Li/C, Be/C
and B/C is presented in terms of relative errors in Fig.~\ref{fig::ErrorBreakdown}.
The sources of uncertainties taken into account are statistics of collected events, 
MC statistics, contamination due to charge mis-identification (\S\ref{sec::ChargeContamination}),
uncertainty on isotopic composition and trigger efficiency uncertainty.

\section{Results and Discussion}  
\label{sec::ResultsAndDiscussion} 

\subsection{The $^{7}$Li/$^{6}$Li Isotopic Ratio} 
\begin{figure*}[!t]
\begin{center}
\epsscale{0.80}
\plotone{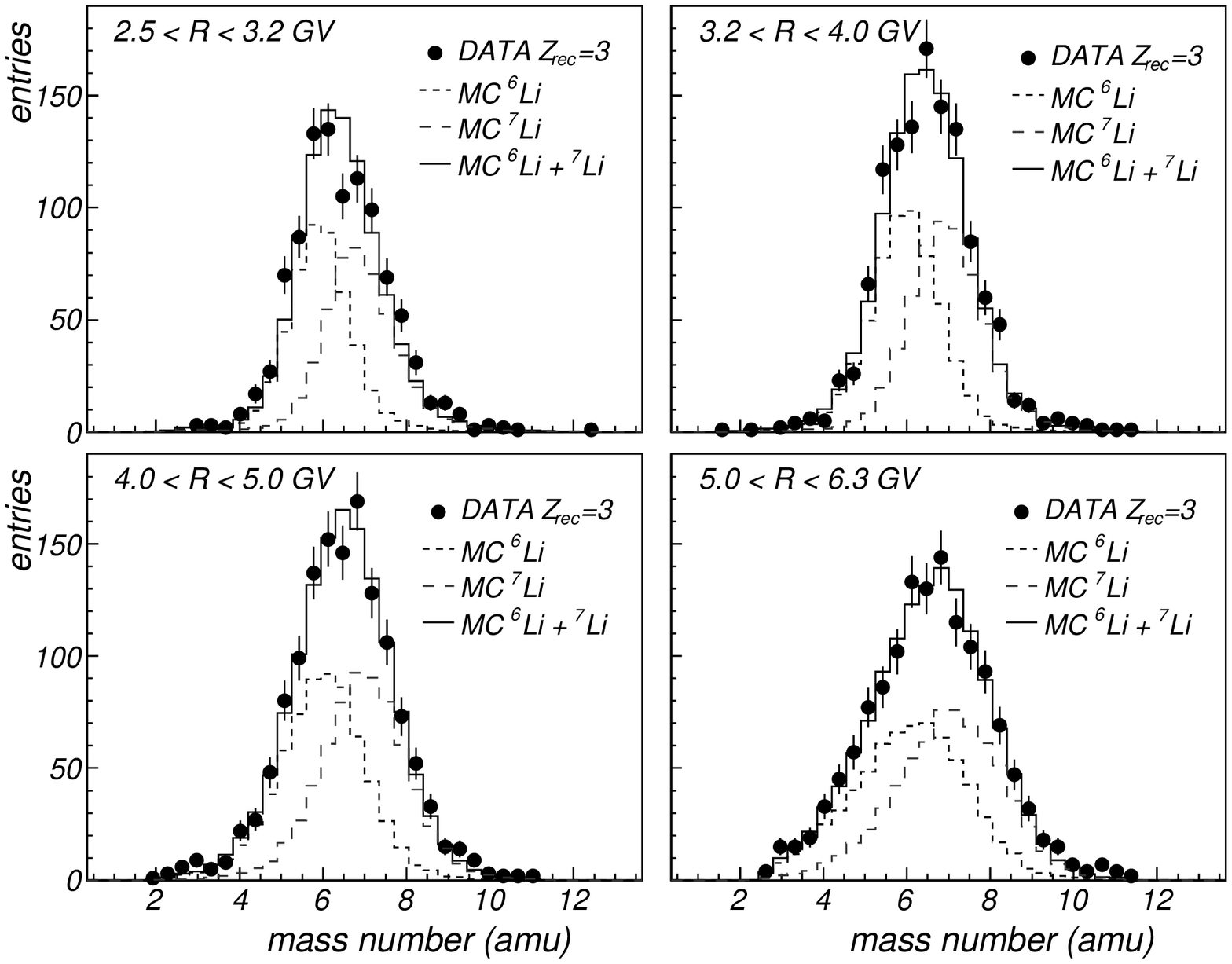} 
\figcaption{
 Lithium mass distributions and best fit composition in four rigidity regions. 
  The amplitudes of the simulated distribution of $^{7}$Li and $^{6}$Li are free parameters.
  \label{fig::LithiumMasses}
}
\end{center}
\end{figure*}
A composition fit technique for determining the isotopic ratio $^{7}$Li/$^{6}$Li 
in four rigidity intervals between 2.5 and 6.3$\,$GV was performed~\citep{Zhou2009}. 
The best fits were obtained by leaving the amplitudes for both isotopes 
free in all four rigidity intervals as shown in Fig.~\ref{fig::LithiumMasses}.
The fits gave unique minima for the $^{7}$Li/$^{6}$Li ratios in each rigidity region, 
providing an average ratio of 1.07~$\pm$~0.16 in the region 2.5$\--$6.3$\,$GV.
\begin{figure}[!h]
\begin{center}
\epsscale{1.00}
\plotone{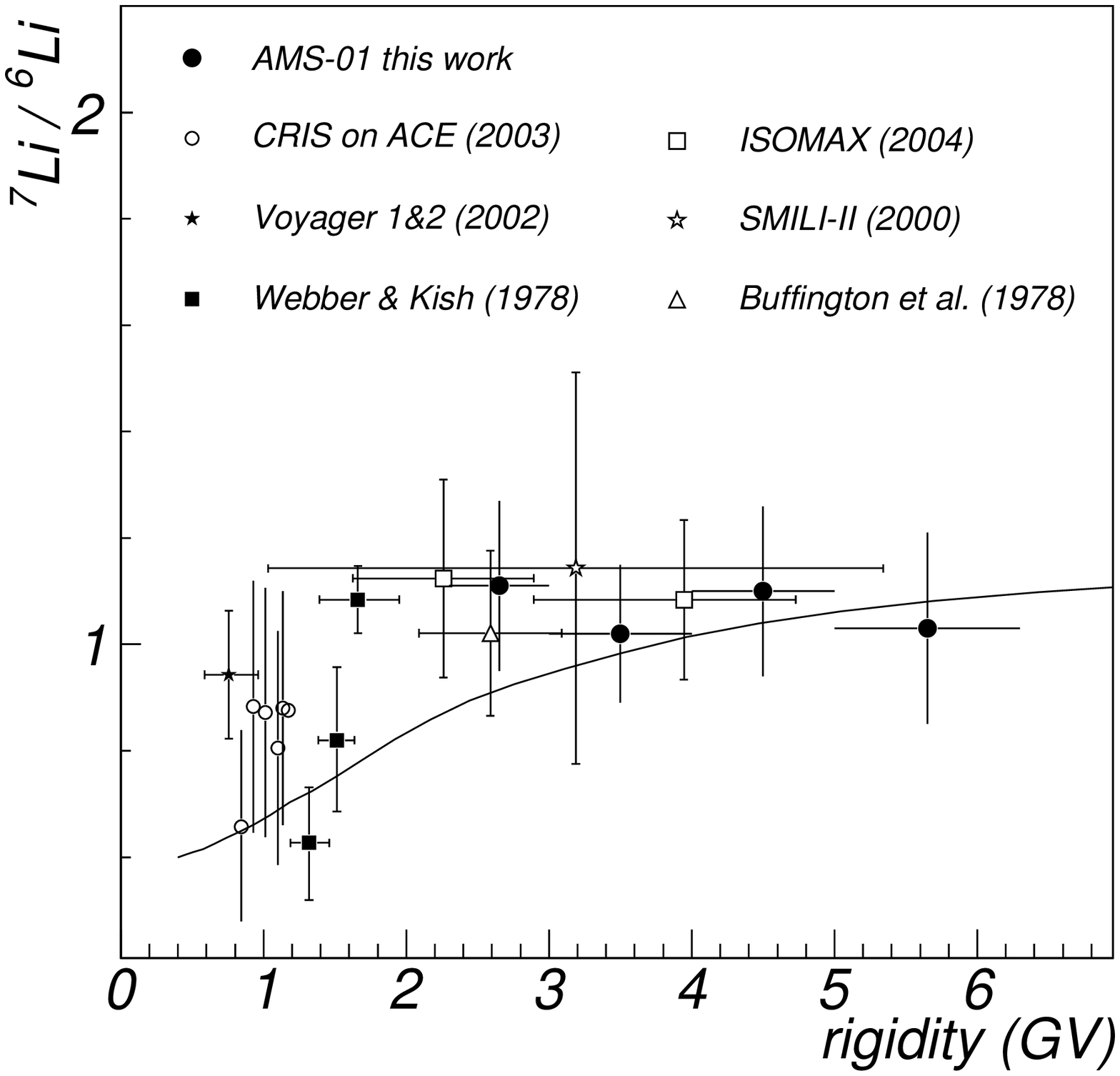}  
\figcaption{
  The rigidity dependence of the ratio $^{7}$Li/$^{6}$Li.
  The other experimental values are converted from kinetic energy to rigidity. 
  The solid line is a diffusion model prediction using \texttt{GALPROP}~(\S\ref{sec::GALPROPcalcs}).  
\label{fig::LithiumIsotopicRatio}
}
\end{center}
\end{figure}
The results are shown in Fig.~\ref{fig::LithiumIsotopicRatio}
with the previous experimental data \citep{DeNolfo2003,Webber2002,Hams2004,Ahlen2000,WebberKish1979,Buffington1978};
the latter are converted into units of rigidity, allowing
a direct comparison with our data. Point values are listed in Table~\ref{tab::LithiumIsotopes}.
Our measurement agrees with the previous data and extends them to higher energies.
In summary, our new measurements of $^{7}$Li/$^{6}$Li extends to 6.3~GV of rigidity 
($\sim$2 GeV~nucleon$^{-1}$ of kinetic energy) with a constant ratio of about equal abundance of both isotopes.
\begin{deluxetable}{ccccccc}[h]
\tablecaption{
  The isotopic ratio $^{7}$Li/$^{6}$Li. 
\label{tab::LithiumIsotopes}}
\tablehead{
  \colhead{Rigidity (GV)} &  \colhead{2.5 -- 3.2} & \colhead{3.2 -- 4.0} & \colhead{4.0 -- 5.0} & \colhead{5.0 -- 6.3} 
}
\startdata
  $^{7}$Li/$^{6}$Li & 1.11 $\pm$ 0.16 & 1.02 $\pm$ 0.13 & 1.10 $\pm$ 0.16 & 1.03 $\pm$ 0.18 \\
    $\chi^{2}$/d.f. & 35.1/35 & 46.1/39 & 41.2/43 & 42.6/43  
\enddata
\end{deluxetable}

\subsection{Secondary to Primary Ratios} 

\begin{figure*}[!t]
\begin{center}
\epsscale{0.85}
\plotone{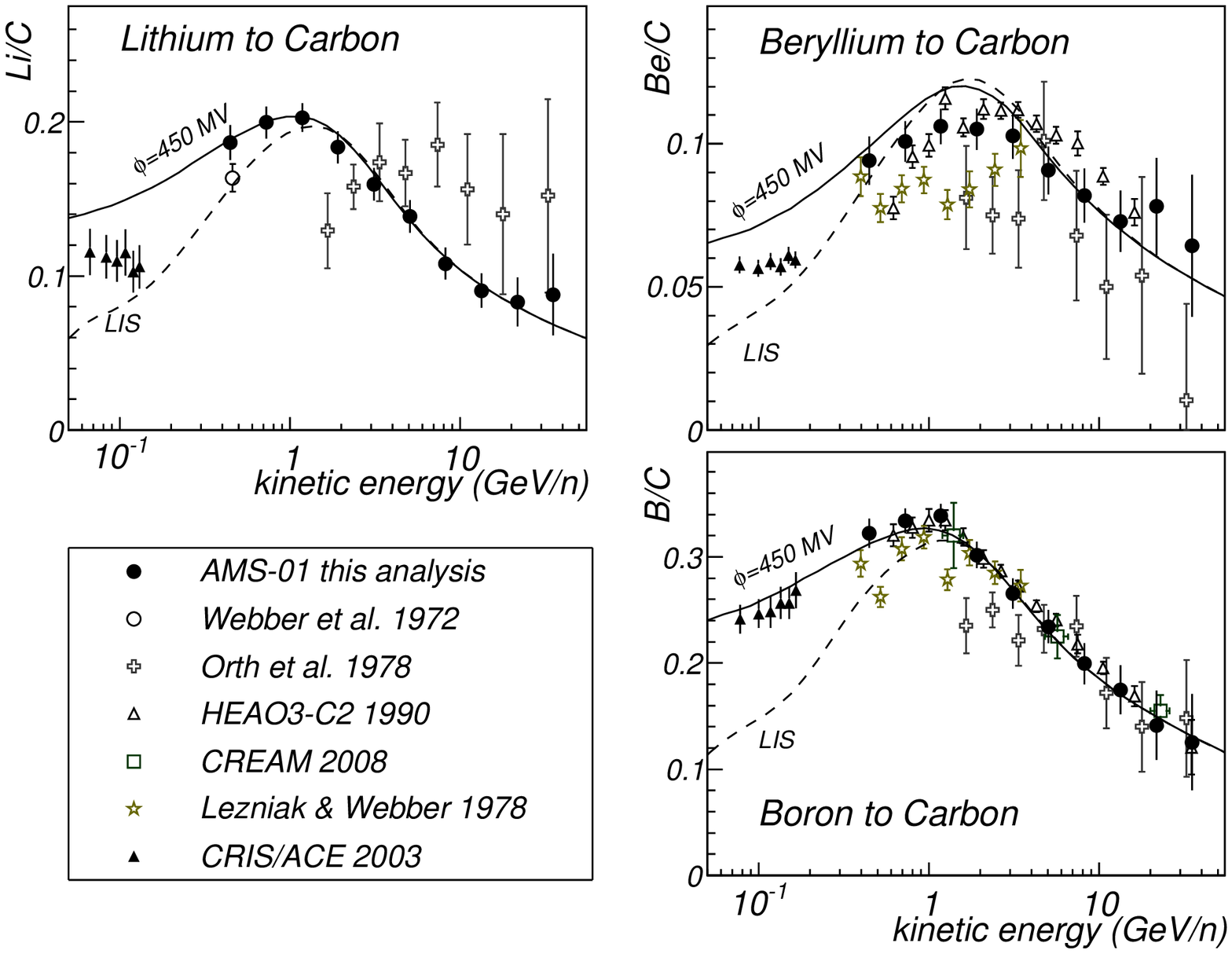} 
\figcaption{
  Results for the secondary to primary ratios Li/C, Be/C and B/C from this work (solid circles)
  and previous measurements \citep{Webber1972, Orth1978, Lezniak1978, HEAO1990, CREAM2008, DeNolfo2003}. 
  Model calculations are also reported, from the \texttt{GALPROP} diffusive-reacceleration model
  predictions for interstellar (LIS) and solar modulated ($\phi$=450 MV) cosmic ray fluxes~\citep{Strong1998}. 
  \label{fig::SecPriRatios}
}
\end{center}
\end{figure*}

Results for the Li/C, Be/C and B/C ratios are presented in Fig.~\ref{fig::SecPriRatios}
with the existing experimental data ~\citep{Webber1972, Orth1978, Lezniak1978, HEAO1990, CREAM2008, DeNolfo2003}. 
The energy range $0.35 \-- 45$\,GeV~nucleon$^{-1}$ is limited by selection inefficiencies 
below $\sim$0.35$\,$GeV~nucleon$^{-1}$ and lack of statistics above $45\,$GeV~nucleon$^{-1}$. 
The error bars in the figure represent the sum in quadrature of 
statistical errors with the systematic uncertainties.
These results are summarized in Table~\ref{tab::SecPriRatios}. 
\begin{deluxetable*}{cc|ccc|ccc|ccc}[!t]
  \tablecaption{ 
    The secondary to primary ratios of Li/C, Be/C and B/C between kinetic energies of $0.35$ and $45\,$GeV/nucleon. 
    The columns indicate: energy intervals $E_{1}$--$E_{2}$, reference values $\langle E \rangle$\tablenotemark{a}, 
    measured ratios, statistical errors and systematic errors.
    \label{tab::SecPriRatios}}
  \tablehead{
    \colhead{ $\Delta$ E } &  \colhead{$\langle$E$\rangle$} & \colhead{Li/C} & \colhead{STAT} & \colhead{SYS} &  \colhead{Be/C} & \colhead{STAT} & \colhead{SYS} & \colhead{B/C} & \colhead{STAT} & \colhead{SYS}
  }
  \startdata 
0.35 -- 0.57  &  0.45  &  0.187  & $\pm$0.009  &  $\pm$0.007  &  0.094  &  $\pm$0.007  &  $\pm$0.005  &  0.323  &  $\pm$0.013  &  $\pm$0.005  \\ 
0.57 -- 0.92  &  0.73  &  0.200  & $\pm$0.007  &  $\pm$0.008  &  0.101  &  $\pm$0.005  &  $\pm$0.005  &  0.334  &  $\pm$0.010  &  $\pm$0.006  \\ 
0.92 -- 1.50  &  1.18  &  0.203  & $\pm$0.006  &  $\pm$0.008  &  0.106  &  $\pm$0.004  &  $\pm$0.004  &  0.339  &  $\pm$0.008  &  $\pm$0.008  \\ 
1.50 -- 2.44  &  1.92  &  0.184  & $\pm$0.004  &  $\pm$0.009  &  0.105  &  $\pm$0.004  &  $\pm$0.006  &  0.302  &  $\pm$0.007  &  $\pm$0.011  \\ 
2.44 -- 3.97  &  3.11  &  0.159  & $\pm$0.004  &  $\pm$0.010  &  0.103  &  $\pm$0.004  &  $\pm$0.007  &  0.265  &  $\pm$0.008  &  $\pm$0.012  \\ 
3.97 -- 6.45  &  5.06  &  0.139  & $\pm$0.005  &  $\pm$0.009  &  0.091  &  $\pm$0.005  &  $\pm$0.007  &  0.234  &  $\pm$0.010  &  $\pm$0.012  \\ 
6.45 -- 10.5  &  8.22  &  0.108  & $\pm$0.006  &  $\pm$0.009  &  0.082  &  $\pm$0.006  &  $\pm$0.007  &  0.199  &  $\pm$0.012  &  $\pm$0.015 \\  
10.5 -- 17.0  &  13.36 &  0.090  & $\pm$0.008  &  $\pm$0.007  &  0.073  &  $\pm$0.009  &  $\pm$0.006  &  0.175  &  $\pm$0.019  &  $\pm$0.013 \\  
17.0 -- 27.7  &  21.72 &  0.083  & $\pm$0.014  &  $\pm$0.007  &  0.078  &  $\pm$0.015  &  $\pm$0.008  &  0.141  &  $\pm$0.030  &  $\pm$0.011 \\  
27.7 -- 45.0  &  35.30 &  0.088  & $\pm$0.026  &  $\pm$0.007  &  0.064  &  $\pm$0.023  &  $\pm$0.010  &  0.126  &  $\pm$0.043  &  $\pm$0.013   
 \enddata  
\tablenotetext{a}{The reference energy $\langle$E$\rangle$ is computed as the geometric mean $\sqrt{E_{1} E_{2}}$.
\\} 
\end{deluxetable*}

Our B/C ratio measurement agrees well with the results from the first flight of 
CREAM in 2004 \citep{CREAM2008} and with the data collected by HEAO-3-C2 \citep{HEAO1990} 
from October 1979 and June 1980. 
The Be/C ratio is consistent, within errors, with the HEAO data, 
but not with balloon data \citet{Orth1978}.
Our Li/C data have unprecedented accuracy in a poorly explored energy region.
In comparison with balloon data from \citet{Orth1978}, 
our data indicate a quite different trend in the high energy part of the Li/C ratio. 

In these ratios, the main progenitors of boron nuclei are primary cosmic rays (CNO).
On the contrary, the abundances of Li and Be depend also on secondary progenitors Be and B
through tertiary contributions like B$\rightarrow$Be, Be$\rightarrow$Li and B$\rightarrow$Li.

However, the observed shapes of the measured ratios Li/C and Be/C suggest 
their suitability in constraining the propagation parameters:   
their decreasing behaviour with increasing energy is a direct consequence of the magnetic diffusion
experienced by their progenitors, while the characteristic peak around $\sim$1$\,$GeV~nucleon$^{-1}$ is 
a strong indicator of low energy phenomena like stochastic reacceleration or 
convective transport with the galactic wind. 

\subsection{Secondary to Secondary Ratios} 

\begin{figure*}[!h]
\begin{center}
\epsscale{0.85}
\plotone{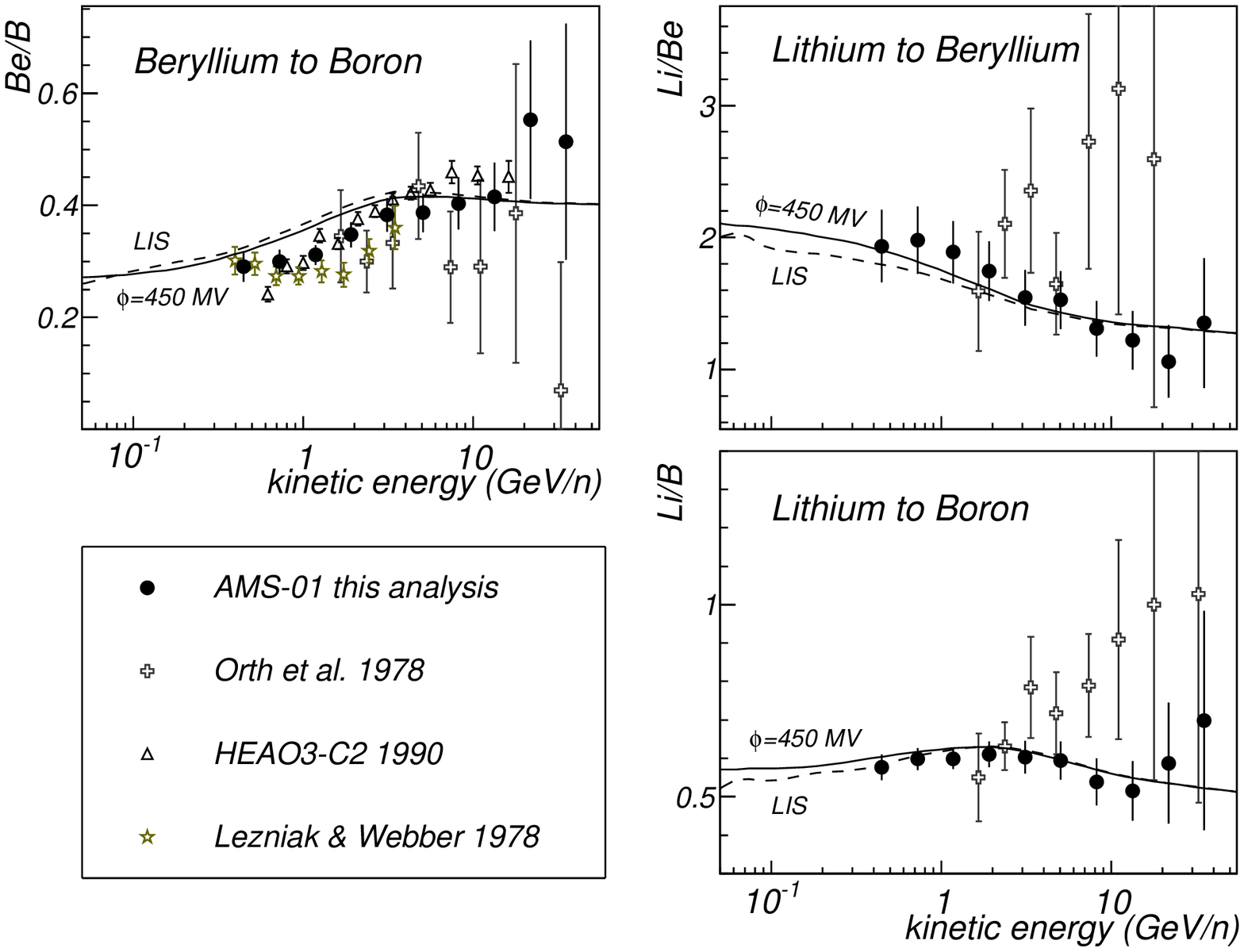} 
\figcaption{
  Secondary to secondary ratios Be/B, Li/Be and Li/B from this work (solid circles)
  and previous measurements \citep{Orth1978, Lezniak1978, HEAO1990}. 
  Model calculations are also reported, from the \texttt{GALPROP} diffusive-reacceleration model
  predictions for interstellar (LIS) and solar modulated ($\phi$=450 MV) cosmic ray fluxes~\citep{Strong1998}. 
  \label{fig::SecSecRatios}
}
\end{center}
\end{figure*}

For sake of completeness, in Fig.~\ref{fig::SecSecRatios} and Table~\ref{tab::SecSecRatios} 
we report the secondary to secondary ratios Li/Be, Li/B and Be/B. 
Though their values can be derived from the secondary to primary ratios of Table~\ref{tab::SecPriRatios}, 
a dedicated analysis of uncertainty was done for these channels. 
These quantities are less sensitive to the propagation parameters,
because all the secondaries have the same origin and undergo similar astrophysical processes.
These ratios are also found to be less influenced by solar modulation. 
Hence, these relative abundances maximize the effects of the nuclear 
aspects of the CR propagation: decays, fragmentations and catastrophic losses.
As pointed out in \citet{WebberSoutol1998}, 
the Be/B ratio is sensitive to the radioactive decay of the beryllium (numerator)
which enriches the boron flux (denominator); thus a precise measurement of the Be/B ratio provides 
constraints on the galactic halo size as well as the $^{10}$Be/$^{9}$Be ratio.
The observed increasing behaviour of the Be/B ratio is due to the relativistic 
lifetime dilation of the unstable Be.
However, the migration $^{10}$Be$\rightarrow$$^{10}$B involves only a small fraction of the elemental flux,
and the present data are still affected by sizeable errors.

\subsection{Propagation Calculations} 
\label{sec::GALPROPcalcs}             

To describe our results, we made use of 
\texttt{GALPROP v50.1p}\footnote{\url{http://galprop.stanford.edu}}, a numerical 
code that incorporates all the astrophysical inputs of the CR galactic transport~\citep{Strong1998}. 
We described the propagation in the framework of the diffusive-reacceleration model, 
that has been very successful in the description of the cosmic ray 
nuclei fluxes~\citep{Moskalenko2002}. 
\texttt{GALPROP} solves the diffusion transport equation for a given source distribution 
and boundary conditions for all the galactic CRs.
In this model, the magnetic diffusion is described in terms of a rigidity dependent 
diffusion coefficient $D = \beta D_{0} \left( R/R_{0} \right)^{\delta}$.
The magnitude of the diffusion, related to the level of hydromagnetic turbulence, 
is controlled by the parameter $D_{0}$ which fixes the normalization of $D(R)$ at the reference rigidity $R_{0}$.
The spectral index $\delta$ is linked to the density of the magnetic irregularities at small scales.
Reacceleration is an energy gain of charged particles due to scattering on 
hydromagnetic waves moving at Alfv\'en speed $V_{a}$ in the ISM.
This process is described as a diffusion in the momentum space and controlled by the parameter $V_{a}$. 
The source spectra $q(R)$ are assumed to be have a pure power law dependence in rigidity, i.e. 
$q(R) \propto R^{-\alpha}$. 
The boundary conditions are expressed by imposing free escape at the galactic halo boundaries,
in particular by the halo height $H$.
Energy losses and catastrophic losses over the interstellar medium are included.
The code makes use of an up to date nuclear reaction network, including
decay rates and fragmentation cross sections for all the relevant channels.
Semiempirical models for cross section calculations are tuned with measured data, 
where available~\citep{MoskalenkoMashnik2003}.
Equilibrium solutions are provided for the local interstellar spectra (LIS) of all the galactic CRs.
In our description, we used a 2D cylindrically symmetric model of the galaxy, 
For the heliospheric propagated fluxes, we adopted the force field description 
of the solar wind~\citep{Gleeson1968}. 
We used $\phi=450\,$MV as modulation parameter, consistent with 1998 June~\citep{Wiedenbeck2005}.
The parameter values of our \texttt{GALPROP} settings are listed in Table~\ref{tab::GalpropSettings}.
\begin{deluxetable}{lcl}[!h]
\tablecaption{
  \texttt{GALPROP} settings.
\label{tab::GalpropSettings}}
\tablehead{
  \colhead{Parameter} & \colhead{Name} & \colhead{Value} 
}
\startdata
diffusion - normalization &  $D_{0}$    &  5.85$\cdot$10$^{28}$ cm$^{2}$s$^{-1}$ \\
diffusion - index  &  $\delta$  &  1/3 \\
reference rigidity &  $R_{0}$    & 4 GV \\
Alfv\'en velocity  &  $V_{a}$    & 30 km s$^{-1}$ \\
injection - index  &  $\alpha$  & 2.43 \\
halo height          &  $H$       & 4 kpc \\
solar modulation   &  $\phi$    & 450 MV  
\enddata
\end{deluxetable}

\begin{deluxetable*}{cc|ccc|ccc|ccc}[!t]
  \tablecaption{ 
    Secondary to secondary ratios of Li/Be, Li/B and Be/B between kinetic energies of $0.35$ and $45$\,GeV/nucleon.
    The columns indicate: energy intervals $E_{1}$--$E_{2}$, reference values $\langle E \rangle$\tablenotemark{a}, 
    measured ratios, statistical errors and systematic errors.
    \label{tab::SecSecRatios}}
  \tablehead{
    \colhead{ $\Delta$ E } &  \colhead{$\langle$E$\rangle$} & \colhead{Li/Be} & \colhead{STAT} & \colhead{SYS} &  \colhead{Li/B} & \colhead{STAT} & \colhead{SYS} & \colhead{Be/B} & \colhead{STAT} & \colhead{SYS}
  }
  \startdata 
0.35 -- 0.57  &  0.45  &  1.935  &  $\pm$0.157  &  $\pm$0.225  &  0.577  &  $\pm$0.030  &  $\pm$0.015  &  0.291  &  $\pm$0.023  &  $\pm$0.014  \\
0.57 -- 0.92  &  0.73  &  1.979  &  $\pm$0.118  &  $\pm$0.227  &  0.598  &  $\pm$0.024  &  $\pm$0.014  &  0.300  &  $\pm$0.017  &  $\pm$0.012  \\
0.92 -- 1.50  &  1.18  &  1.889  &  $\pm$0.085  &  $\pm$0.221  &  0.599  &  $\pm$0.020  &  $\pm$0.017  &  0.311  &  $\pm$0.013  &  $\pm$0.012  \\ 
1.50 -- 2.44  &  1.92  &  1.747  &  $\pm$0.066  &  $\pm$0.216  &  0.610  &  $\pm$0.020  &  $\pm$0.027  &  0.348  &  $\pm$0.013  &  $\pm$0.019  \\ 
2.44 -- 3.97  &  3.11  &  1.544  &  $\pm$0.065  &  $\pm$0.202  &  0.603  &  $\pm$0.024  &  $\pm$0.035  &  0.384  &  $\pm$0.015  &  $\pm$0.026  \\ 
3.97 -- 6.45  &  5.06  &  1.528  &  $\pm$0.084  &  $\pm$0.203  &  0.594  &  $\pm$0.032  &  $\pm$0.037  &  0.387  &  $\pm$0.022  &  $\pm$0.027  \\ 
6.45 -- 10.5  &  8.22  &  1.310  &  $\pm$0.105  &  $\pm$0.183  &  0.539  &  $\pm$0.045  &  $\pm$0.042  &  0.404  &  $\pm$0.033  &  $\pm$0.033  \\  
10.5 -- 17.0  &  13.36  &  1.223  &  $\pm$0.157  &  $\pm$0.155  &  0.515  &  $\pm$0.069  &  $\pm$0.036  &  0.416  &  $\pm$0.057  &  $\pm$0.022  \\  
17.0 -- 27.7  &  21.72  &  1.062  &  $\pm$0.238  &  $\pm$0.135  &  0.588  &  $\pm$0.151  &  $\pm$0.045  &  0.553  &  $\pm$0.139  &  $\pm$0.023  \\ 
27.7 -- 45.0  &  35.30  &  1.352  &  $\pm$0.442  &  $\pm$0.218  &  0.699  &  $\pm$0.280  &  $\pm$0.056  &  0.514  &  $\pm$0.202  &  $\pm$0.060  
\enddata  
\tablenotetext{a}{The reference energy $\langle$E$\rangle$ is computed as the geometric mean $\sqrt{E_{1} E_{2}}$.
\\} 
\end{deluxetable*}
In Fig.~\ref{fig::SecPriRatios} and \ref{fig::SecSecRatios} our data are compared with the model calculation in 
the heliosphere (solid lines). The LIS ratios (dashed lines) are shown for reference. 
Our results for the B/C ratio, the Li/C ratio and the $^{7}$Li/$^{6}$Li isotopic fraction 
of Fig.~\ref{fig::LithiumIsotopicRatio} are described quite well within the uncertainties.
It is difficult, however, to accomodate the Li and B description with the Be/C ratio 
by only means of astrophysical parameters.
Beryllium appears to be overproduced in the model by a factor $\sim$10$\--$15$\,$\%. 
This discrepancy is also apparent in the previous measurements.
The ratios Be/B and Li/Be of Fig.~\ref{fig::SecSecRatios} also indicate this feature,
whereas the Li/B ratio is well described by our \texttt{GALPROP} tune. 

It is worth noting that the production of Li and Be involves very complex reaction chains,
due to the multichannel character of the CR fragmentation.
Moreover, cross sections for the tertiary processes LiBeB + ISM $\rightarrow$ LiBe
are less well known than for the interactions CNO + ISM $\rightarrow$ LiBeB.
Spallation contributions like $^{11}$B$\rightarrow$$^{9}$Be are measured at only a few energy points.
Other channels like $^{10}$B$\rightarrow$$^{9}$Be, $^{14,15}$N$\rightarrow$$^{9}$Be or
$^{10,11}$B$\rightarrow$$^{7}$Be are not measured at all and rely on 
extrapolated parametrizations~\citep{MoskalenkoMashnik2003}. 
Spallation processes involving interstellar helium are even less well understood.
The lack of cross section measurements limits the model predictions 
to uncertainties of $\sim$10--20$\,$\% in the Li/C and Be/C ratios~\citep{DeNolfo2006}. 
Understanding fragmentation is a key factor in establishing
final conclusions concerning cosmic ray propagation.

\section{Conclusions}
We have presented a new measurement of light CR nuclei composition
in the energy range from $0.35 \-- 45\,$GeV~nucleon$^{-1}$ with the AMS-01 experiment.
The isotopic ratio $^{7}$Li/$^{6}$Li has been measured between 2.5 and 6.3 GV of rigidity.
This work is aimed at investigating the origin and the physical 
properties of the galactic cosmic rays. 

The study of high charged ions with AMS-01 required the
development of an improved charge identification algorithm.
A thorough analysis has been made to understand the CR interactions
in the spectrometer, its instrumental response and the orbital environment. 
Our results for the B/C ratio agree well with data collected from
the HEAO-3-C2 experiment and with the more recent measurement 
from CREAM.  
The Li/C ratio has been measured with unprecedent accuracy.
A 10$\--$15$\,$\% overproduction of Be is found in the model predictions,
that describe well Li and B. 
This is consistent with the lack of cross section measurements, that 
limits the model predictions up to uncertainties of $\sim$20$\,$\% for light-nuclei.

The astrophysical interest of light nuclei cosmic ray fluxes measurements,
both for propagation studies and in the search for exotic phenomenon,
has been well established in the CR community.
We expect, with AMS-02, to measure them
with high precision over wide energy ranges in the near future.

\section{Acknowledgments} 
The support of INFN, Italy, ETH-Zurich, the University of
Geneva, the Chinese Academy of Sciences, Academia Sinica
and National Central University, Taiwan, the RWTH Aachen,
Germany, the University of Turku, the University of Technology
of Helsinki, Finland, the US DOE and MIT, CIEMAT,
Spain, LIP, Portugal and IN2P3, France, is gratefully acknowledged.
The success of the first AMS mission is due to many individuals
and organizations outside of the collaboration. The support
of NASA was vital in the inception, development and operation
of the experiment. Support from the Max-Planck Institute for Extraterrestrial Physics, 
from the space agencies of Germany (DLR), Italy (ASI), France (CNES) 
and China and from CSIST, Taiwan played important roles in the success of AMS.\\


\end{document}